\documentclass[conference]{IEEEtran}
\IEEEoverridecommandlockouts
\usepackage{amsmath,amssymb,amsfonts}
\usepackage{graphicx}
\usepackage{textcomp}
\usepackage{xcolor}
\def\BibTeX{{\rm B\kern-.05em{\sc i\kern-.025em b}\kern-.08em
    T\kern-.1667em\lower.7ex\hbox{E}\kern-.125emX}}

\usepackage{comment}

\usepackage{nohyperref}

\definecolor{matplotlib0}{HTML}{1f77b4}
\definecolor{matplotlib1}{HTML}{d62728}
\definecolor{matplotlib2}{HTML}{2ca02c}
\definecolor{matplotlib3}{HTML}{ff7f0e}
\definecolor{matplotlib4}{HTML}{9467bd}
\definecolor{matplotlib5}{HTML}{8c564b}
\definecolor{matplotlib6}{HTML}{e377c2}
\definecolor{matplotlib7}{HTML}{7f7f7f}
\definecolor{matplotlib8}{HTML}{bcbd22}
\definecolor{matplotlib9}{HTML}{17becf}

\usepackage{mathtools} 

\usepackage{subcaption}

\usepackage{booktabs}
\usepackage{multirow}
\usepackage{colortbl}
\usepackage{tablefootnote}
\usepackage{threeparttable}
\usepackage{makecell}

\usepackage[acronym, style=super, nonumberlist]{glossaries}

\newcommand{\wolf}[0]{{Mr.\,Wolf}}
\newcommand{\riscy}[0]{\textsc{RI5CY}}
\newcommand{\eegnet}[0]{\textsc{EEGNet}}
\newcommand{\pulp}[0]{\textsc{PULP}}

\newcommand{\riscv}[0]{\textsc{RISC-V}}
\newcommand{\xpulp}[0]{\textsc{XpulpV2}}

\usepackage{tikz}
\usetikzlibrary{shapes, arrows}
\tikzset{>=latex}

\usepackage{setspace} 

\tikzstyle{block} = [draw, thick, rectangle, minimum height=0.75cm, minimum width=0.75cm]
\tikzstyle{sum} = [draw, fill=white, circle, node distance=1cm, thick]
\tikzstyle{gain} = [
  regular polygon,
  regular polygon sides=3,
  draw,
  thick,
  fill=white,
  text width=0.3em,
  inner sep=0.3mm,
  outer sep=0mm,
  shape border rotate=-90
]
\tikzstyle{revgain} = [
  regular polygon,
  regular polygon sides=3,
  regular polygon rotate=90,
  draw,
  thick,
  fill=white,
  text width=0.3em,
  inner sep=0.3mm,
  outer sep=0mm,
]

\usepackage{pgfplots}
\usepgfplotslibrary{fillbetween}
\usepgfplotslibrary{colormaps}
\pgfplotsset{
            compat=1.16,
            label style={font=\footnotesize},
            tick label style={font=\scriptsize},
            legend style={font=\scriptsize}
            }

\pgfplotscreateplotcyclelist{matplotlib}{
  {matplotlib0},
  {matplotlib1},
  {matplotlib2},
  {matplotlib3},
  {matplotlib4},
  {matplotlib5},
  {matplotlib6},
  {matplotlib7},
  {matplotlib8},
  {matplotlib9}
}

\pgfplotsset{every axis/.append style={
    cycle list name=matplotlib,
}}

\usepackage{listings}

\definecolor{code_default}{HTML}{000000}
\definecolor{code_keyword}{HTML}{AC4142}
\definecolor{code_identifier}{HTML}{D28445}

\lstdefinelanguage{RISCV}{
  sensitive=false,
  morecomment=[l]{//},
  alsoletter={.},
  morekeywords=[1]{
    lp.setup, mv, lw, p.lw, sw, p.sw, pv.sdotsp.b, pv.shuffle2.b, p.subNR, p.addNR
  },
  morekeywords=[2]{
    zero, ra, sp, gp, tp, t0, t1, t2, t3, t4, t5, t6, s0, s1, a0, a1, a2, a3, a4, a5, a6, a7, a8, a9, a10, a11,
  },
  morestring=[b]",
  morestring=[b]',
}[strings, comments, keywords]

\lstdefinestyle{RISCV_STYLE}{
  language=RISCV,
  numbers=none,
  basicstyle=\scriptsize\ttfamily\color{code_default},
  keywordstyle=[1]\color{matplotlib0},
  keywordstyle=[2]\color{matplotlib1},
  float,
  captionpos=b,
  belowskip=-0.5cm
}

\usepackage{algorithm}
\usepackage{algpseudocode}
\usepackage{float}
\newfloat{algorithm}{t}{top}

\newacronym{simd}{SIMD}{Single Instruction, Multiple Data}
\newacronym{elu}{ELU}{Exponential Linear Unit}
\newacronym{relu}{ReLU}{Rectified Linear Unit}
\newacronym{rpr}{RPR}{Random Partition Relaxation}
\newacronym{mac}{MAC}{Multiply Accumulate}
\newacronym{dma}{DMA}{Direct Memory Access}
\newacronym{cnn}{CNN}{Convolutional Neural Network}
\newacronym{dnn}{DNN}{Convolutional Neural Network}
\newacronym{bmi}{BMI}{Brain--Machine Interface}
\newacronym{bci}{BCI}{Brain--Computer Interface}
\newacronym{smr}{SMR}{Sensory Motor Rythms}
\newacronym{eeg}{EEG}{Electroencephalography}
\newacronym{svm}{SVM}{Support Vector Machine}
\newacronym{svd}{SVD}{Singular Value Decomposition}
\newacronym{evd}{EVD}{Eigendecomposition}
\newacronym{iir}{IIR}{Infinite Impulse Response}
\newacronym{fir}{FIR}{Finite Impulse Response}
\newacronym{fc}{FC}{Fabric Controller}
\newacronym{nn}{NN}{Neural Network}
\newacronym{mrc}{MRC}{Multispectral Riemannian Classifier}
\newacronym{flop}{FLOP}{Floating Point Operation}
\newacronym{sos}{SOS}{Second-Order Section}
\newacronym{ipc}{IPC}{Instructions per Cycle}
\newacronym{tcdm}{TCDM}{Tightly Coupled Data Memory}
\newacronym{fpu}{FPU}{Floating Point Unit}
\newacronym{fma}{FMA}{Fused Multiply Add}
\newacronym{alu}{ALU}{Arithmetic Logic Unit}
\newacronym{dsp}{DSP}{Digital Signal Processing}
\newacronym{gpu}{GPU}{Graphics Processing Unit}
\newacronym{soc}{SoC}{System-on-Chip}
\newacronym{mi}{MI}{Motor-Imagery}
\newacronym{csp}{CSP}{Common Spatial Patterns}
\newacronym{fbcsp}{FBCSP}{Filter-Bank \acrlong{csp}}
\newacronym{pulp}{PULP}{parallel ultra-low power}
\newacronym{soa}{SoA}{state-of-the-art}
\newacronym{bn}{BN}{Batch Normalization}
\newacronym{isa}{ISA}{Instruction Set Architecture}
\newacronym{mmm}{MMM}{matrix-matrix multiplication}
\newacronym{mcu}{MCU}{microcontroller unit}
\newacronym{ssvep}{SSVEP}{steady-state visual evoked potential}

\renewcommand{\baselinestretch}{0.905}

\newcommand{\new}[1]{{\color{black}#1}}

\usepackage{fancyhdr}
\fancypagestyle{mahmood}{%
  \fancyhf{} 
  
  \fancyhead[C]{\footnotesize \textcopyright 2021 IEEE. Personal use of this material is permitted.  Permission from IEEE must be obtained for all other uses, in any current or future media, including reprinting/republishing this material for advertising or promotional purposes, creating new collective works, for resale or redistribution to servers or lists, or reuse of any copyrighted component of this work in other works.}
}%
\makeatletter
\let\ps@IEEEtitlepagestyle\ps@mahmood
\makeatother

\begin{document}

\title{Mixed-Precision Quantization and Parallel Implementation of Multispectral Riemannian Classification for Brain--Machine Interfaces\vspace{-0.3cm}
}


\author{\IEEEauthorblockN{
    Xiaying Wang\IEEEauthorrefmark{1}, 
    Tibor Schneider\IEEEauthorrefmark{1},
    Michael Hersche\IEEEauthorrefmark{1}, 
    Lukas Cavigelli\IEEEauthorrefmark{2},
    Luca Benini\IEEEauthorrefmark{1}\IEEEauthorrefmark{3}}
    \IEEEauthorblockA{\\[-2mm]\IEEEauthorrefmark{1}ETH Zürich, D-ITET, Switzerland \hspace{3.2mm}\IEEEauthorrefmark{3}University of Bologna, DEI, Italy \hspace{3.2mm}
    \IEEEauthorrefmark{2}Huawei Technologies, Zurich RC, Switzerland\vspace{-0.3cm}}
    \thanks{Corresponding emails: \{xiaywang, herschmi\}@iis.ee.ethz.ch}
    \vspace{-0.3cm}
    }


\maketitle

\begin{abstract}
\new{
With \acrfull{mi} \acrfullpl{bmi} we may control machines by merely thinking of performing a motor action. Practical use cases require a wearable solution where the classification of the brain signals is done locally near the sensor using machine learning models embedded on energy-efficient \acrfullpl{mcu}, for assured privacy, user comfort, and long-term usage. 
In this work, we provide practical insights on the accuracy-cost trade-off for embedded \acrshort{bmi} solutions. Our proposed Multispectral Riemannian Classifier
reaches 75.1\% accuracy on 4-class \acrshort{mi} task. We further scale down the model by quantizing it to mixed-precision representations with a minimal accuracy loss of 1\%, which is still 3.2\% more accurate than the state-of-the-art embedded convolutional neural network. We implement the model on a low-power \acrshort{mcu} with parallel processing units taking only 33.39\,ms and consuming 1.304\,mJ per classification.


}
\end{abstract}


\section{Introduction}\label{ch:introduction}

\gls{mi} \glspl{bmi} use \gls{eeg} signals recorded from the brain to decode a movement imagined by the subject. The decoded information can be used to control an external device, such as a drone~\cite{Koizumi2018_BMI_visual} or a wheelchair~\cite{Yu2017Self-PacedPotential, xiong2019low}, or for stroke rehabilitation~\cite{Frolov2017Post-strokeTrial}. It is especially useful for individuals with physical disabilities to regain independence~\cite{Frolov2017Post-strokeTrial,Kobayashi2018BCI-basedEEG-short}. 
However, the high variability across subjects and among different recording sessions poses big challenges to an accurate \gls{mi}-\gls{bmi}. Moreover, recording and labelling \gls{eeg} data is expensive, time consuming, and prone to errors, resulting in scarce amounts of data available for training complex models with large numbers of parameters. In fact, many studies using \glspl{cnn} acknowledge the fact that overfitting is the biggest issue for these types of models~\cite{leon2020,wu2019,Schirrmeister2017DeepVisualization}.

On the other hand, successful methods have been proposed to extract discriminative, domain-specific features from \gls{eeg} signals. The well-known \gls{csp} learns spatial filters that discern between different \gls{mi} tasks~\cite{Lotte2011RegularizingAlgorithms}. An improved algorithm, called Filter-Bank \gls{csp}, that accounts for multiple frequency bands achieved better accuracy~\cite{KaiKengAng2008FilterInterface}. More recent studies have proposed Riemannian methods to extract more comprehensive features also in absence of labeled data~\cite{NGUYEN20181871}.
The unsupervised feature calibration enables online adaptation of the classifier to combat the large inter-session variance in MI-BMIs~\cite{lotteAdaptive}. 
So far, these methods are believed to be the most promising feature extractors for several kinds of \gls{bmi} paradigms~\cite{congedo2017RiemannianReview,Yger2017RiemannianReview,riemannianSSVEPreview}. 

Traditional \gls{bmi} systems adopt offline, remote processing of the sensor data, raising concerns over data privacy, latency, high energy consumption, and battery lifetime. A promising solution is to bring the processing near the sensor, i.e. on the body of the user, using low-power low-cost \glspl{mcu}, allowing the data to be processed locally~\cite{wang2019fann}. 
However, these devices suffer from limited on-board resources in terms of memory and computational capabilities. Hence, researching compact yet accurate algorithms~\cite{wu2019,Lawhern2018EEGNet:Interfaces} and designing low-power processors with high capabilities~\cite{pullini2019wolf} has become an emerging trend. 
Most of the \gls{mi}-\gls{bmi} models, particularly \glspl{cnn}, are too demanding for low-power \glspl{mcu}~\cite{Ingolfsson2020}. 
TPCT~\cite{li_novel_2020} reached the \gls{soa} accuracy of 88.87\% on the BCI Competition IV-2a dataset~\cite{Brunner2008BCIA}. The model consists of around 7.78\,M parameters. 
Other similar \glspl{cnn} reach 81.1\% with 240\,k parameters~\cite{zhao_improvement_2017} or 75.8\% with 155\,k parameters~\cite{wu_parallel_2019}. 
A notable exception is \eegnet{}~\cite{Lawhern2018EEGNet:Interfaces} with only few thousands parameters, i.e. three orders of magnitude less demanding, but still achieving around 70\% accuracy on 4-class \gls{mi} classification.
By virtue of its compactness, 
it has been successfully quantized with Q-\eegnet{}~\cite{Schneider2020} and implemented on a low-power \gls{soc} based on RISC-V called \wolf{}~\cite{pullini2019wolf}. It has proven to be three orders of magnitude more energy efficient than an implementation on commercially available \glspl{mcu} based on ARM \mbox{Cortex-M} architecture~\cite{Wang2020Memea}, making it the \gls{soa} embedded \gls{cnn} in terms of energy efficiency, compact model size yet accurate performance.
Another effort for embedded \gls{bmi} has been made by Belwafi et al.~\cite{Belwafi2018AnSystems} implementing a \gls{csp}-based classifier on a FPGA device. 
The multispectral and multiscale Riemannian classifiers proposed in~\cite{Hersche2018FastFeatures,Yang2020ieeeaccess} outperform both \eegnet{} and \gls{csp}-based models by around 5\% and 2\% higher accuracy, respectively. However, their proposed models are still very challenging for embedded deployment on low-power resource-constrained \glspl{mcu} due to large memory footprint and high computational complexity.

For the first time in literature, we propose an embedded \gls{mi}-\gls{bmi} based on a Riemannian classifier~\cite{Hersche2018FastFeatures}. The main contributions of this paper are:
(a)  We tailor the model for better embedded deployment by reducing its size and complexity, i.e., the number of frequency bands and temporal windows, while at the same time keeping comparable classification accuracy by introducing regularization (75.1\% ours vs. 75.5\%~\cite{Hersche2018FastFeatures}).
(b) We further quantize the \gls{mrc} from full precision (32-bit float) to a mixture of precisions with 8-, 16-, 32-bit fixed- and floating-point representations, to maximize efficiency on low-power \glspl{mcu} by enabling the use of fixed-point SIMD instructions while maintaining a minimal accuracy loss. The quantization yields 1\% accuracy drop which is still 3.2\% more accurate than the embedded \gls{cnn}-based \eegnet{} (74.1\% ours vs. 70.9\%~\cite{Schneider2020}).
(c) We efficiently implement the mixed-precision model on \wolf{} by exploiting the underlying hardware architecture, i.e., custom \gls{isa} extensions and concurrent execution on multiple cores and measure the performance on-board. Experimental measurements show that the proposed model takes only 33.30\,ms and consumes 1.304\,mJ per inference.
(d) Our work provides a practical accuracy-cost trade-off between \gls{mrc}, a discriminative feature-based approach, and \eegnet{} as a \gls{cnn}-based approach, supported by an actual implementation and measurement results. Being the first embedded implementation of Riemannian covariance kernels and the most accurate embedded \gls{mi}-\gls{bmi}, it opens the path for other \gls{bmi} paradigms deploying Riemmanian methods, e.g., steady-state visual evoked potential~\cite{riemannianSSVEPreview} and P300~\cite{TLriemannianP300}.
Finally, we release open-source code\footnote{https://github.com/pulp-platform/multispectral-riemannian}. 

\section{Design and Quantization}\label{ch:riemann}

\begin{figure*}[t]
  \centering
  \begin{tikzpicture}[scale=0.7, transform shape, node distance=3.25cm]
    \node (input) {$\mathbf{X}$};
    \node[block, right of=input, yshift=+0.8cm, xshift=-1cm] (iir1) {IIR $b_1$};
    \node[right of=input, yshift=0.1cm, xshift=-1cm] {$\vdots$};
    \node[block, right of=input, yshift=-0.8cm, xshift=-1cm] (iir2) {IIR $b_f$};
    \node[block, right of=iir1, xshift=-0.4cm] (cm1) {$\mathbf{X}_1 \mathbf{X}_1^T + \mathbf{I}\rho$};
    \node[right of=iir1, yshift=-0.7cm, xshift=-0.4cm] {$\vdots$};
    \node[block, right of=iir2, xshift=-0.4cm] (cm2) {$\mathbf{X}_f \mathbf{X}_f^T + \mathbf{I}\rho$};
    \node[block, right of=cm1, xshift=0.2cm] (wh1) {$\mathbf{C}_{\text{ref},1}^{-1/2} \mathbf{C} \mathbf{C}_{\text{ref},1}^{-1/2}$};
    \node[right of=cm1, yshift=-0.7cm, xshift=0.2cm] {$\vdots$};
    \node[block, right of=cm2, xshift=0.2cm] (wh2) {$\mathbf{C}_{\text{ref},f}^{-1/2} \mathbf{C}_f \mathbf{C}_{\text{ref},f}^{-1/2}$};
    \node[block, right of=wh1, xshift=0.2cm] (logm1) {$\text{logm}(\mathbf{W}_1)$};
    \node[right of=wh1, yshift=-0.7cm, xshift=0.2cm] {$\vdots$};
    \node[block, right of=wh2, xshift=0.2cm] (logm2) {$\text{logm}(\mathbf{W}_f)$};
    \node[block, right of=logm1, xshift=-0.4cm] (diag1) {$\text{vect}(\mathbf{L}_1)$};
    \node[right of=logm1, yshift=-0.7cm, xshift=-0.4cm] {$\vdots$};
    \node[block, right of=logm2, xshift=-0.4cm] (diag2) {$\text{vect}(\mathbf{L}_f)$};

    \draw (iir1) node[above, yshift=0.5cm, align=center] {\large Filter};
    \draw (cm1) node[above, yshift=0.5cm, text width=2cm, align=center] {\large Covariance Matrix};
    \draw (wh1) node[above, yshift=0.5cm, align=center] {\large Whitening};
    \draw (logm1) node[above, yshift=0.5cm, text width=2cm, align=center] {\large Matrix Logarithm};

    \draw[thick] (16.50, 1.175) rectangle node[pos=0.5, rotate=90](svm){SVM} (17.25, -1.175);
    \draw[thick] (18.10, 0.75) circle (0.1cm);
    \draw[thick] (18.10, 0.25) circle (0.1cm);
    \draw[thick] (18.10, -0.25) circle (0.1cm);
    \draw[thick] (18.10, -0.75) circle (0.1cm);
    \draw[thick] (18.50, -0.75) node[right, rotate=90] {4 classes};

    \draw[->] (input) -- ++(0.75, 0) |- (iir1);
    \draw[->] (input) -- ++(0.75, 0) |- (iir2);
    \path[->] (iir1) edge node[above] {$\mathbf{X}_1$} (cm1);
    \path[->] (iir2) edge node[above] {$\mathbf{X}_f$} (cm2);
    \path[->] (cm1) edge node[above] {$\mathbf{C}_1$} (wh1);
    \path[->] (cm2) edge node[above] {$\mathbf{C}_f$} (wh2);
    \path[->] (wh1) edge node[above] {$\mathbf{W}_1$} (logm1);
    \path[->] (wh2) edge node[above] {$\mathbf{W}_f$} (logm2);
    \path[->] (logm1) edge node[above] {$\mathbf{L}_1$} (diag1);
    \path[->] (logm2) edge node[above] {$\mathbf{L}_f$} (diag2);
    \path[->] (diag1) edge (16.50, 0.8);
    \path[->] (diag2) edge (16.50, -0.8);
    \path[->] (17.25, 0.75) edge (18, 0.75);
    \path[->] (17.25, 0.25) edge (18, 0.25);
    \path[->] (17.25, -0.25) edge (18, -0.25);
    \path[->] (17.25, -0.75) edge (18, -0.75);
    
    \node (rect) at (5.65, 0) [draw,thick,minimum width=8.75cm,minimum height=2.6cm, dashed, blue] {};
    \node (rect) at (5.6, 0.8) [draw=none,thick,minimum width=8.5cm,minimum height=0.85cm, fill=blue, opacity=0.15] {};
    
    \node (rect) at (13.35, 0) [draw,thick,minimum width=5cm,minimum height=2.6cm, dashdotted, red] {};
    \node (rect) at (13.35, 0.6) [draw=none,thick,minimum width=4.95cm,minimum height=1.25cm, fill=red, opacity=0.15] {};
    
    \node (rect) at (17.25, 0) [draw,thick,minimum width=2.15cm,minimum height=2.6cm, dotted, green] {};
    \node (rect) at (17.25, 0) [draw=none,thick,minimum width=2cm,minimum height=2.45cm, fill=green, opacity=0.15] {};
  \end{tikzpicture}
  \caption{\acrlong{mrc} with $n=18$ frequency bands and one time window.}\label{fig:background:riemannian}
  \vspace{-0.3cm}
\end{figure*}
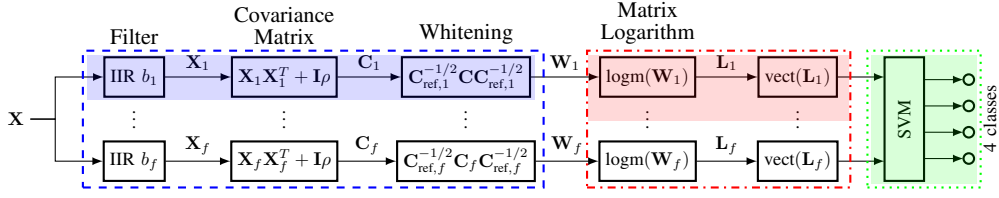

\begin{figure*}[t]
  \centering
  \begin{tikzpicture}[scale=0.7, transform shape, node distance=2.25cm]
    \node (input) {$\tilde{\mathbf{X}}$};
    \node[block, right of=input, xshift=-0.2cm](iir){$\text{IIR}_{Q12, Q16}$};
    \node[block, right of=iir, xshift=0.5cm](covmat){Covmat$_{Q16}$};
    \node[block, right of=covmat, xshift=0.8cm](white){Whitening$_{Q11}$};
    \node[block, right of=white, xshift=0.8cm](deq){Dequantize};
    \node[block, right of=deq, xshift=0.4cm](logm){$\text{logm}_{F32}$};
    \node[block, right of=logm, xshift=0.4cm](quant){Requantize};
    \node[block, right of=quant, xshift=0.2cm](vec){vect$_{Q8}$};
    \node[block, right of=vec](svm){SVM$_{Q8}$};
    \node[draw, circle, thick, right of=svm, xshift=-0.5cm, inner sep=0.065cm] (out) {};

    \draw[->] (input) -- node[above]{Q8} (iir);
    \draw[->] (iir) -- node[above]{Q8} (covmat);
    \draw[->] (covmat) -- node[above]{Q16} (white);
    \draw[->] (white) -- node[above]{Q32} (deq);
    \draw[->] (deq) -- node[above]{F32} (logm);
    \draw[->] (logm) -- node[above]{F32} (quant);
    \draw[->] (quant) -- node[above]{Q8} (vec);
    \draw[->] (vec) -- node[above]{Q8} (svm);
    \draw[->] (svm) -- node[above]{Q32} (out);
  \end{tikzpicture}
  \caption{Quantized \gls{mrc} of a single frequency band, showing the representation of each intermediate signal.}\label{fig:riemann:feature_extraction_quant}
  \vspace{-0.3cm}
\end{figure*}
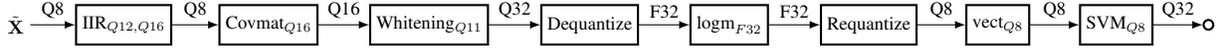

\gls{mrc}~\cite{Hersche2018FastFeatures} consists of a non-linear feature extraction applying the Riemannian covariance method~\cite{yger2016riemannian} on multiple frequency bands and temporal windows, followed by a linear \gls{svm}, depicted in Fig.~\ref{fig:background:riemannian}.
First, the input data is filtered using $f$ different \gls{iir} bandpass filters. Then, the covariance matrix is estimated and regularized with the parameter $\rho$.
The next block, called \emph{Whitening}, multiplies from the left and right with a reference matrix $\mathbf{C}_{\text{ref},k}^{-1/2}$, that is computed for each frequency band $k$ independently during training.
Afterwards, the matrix logarithm is computed with the help of \gls{evd}.
Then, the function $\text{vect}(\mathbf{L}_k)$ vectorizes the symmetric matrix $\mathbf{L}_k$ by concatenating the diagonal values and the upper right non-diagonal elements. To preserve the norm, the off-diagonal elements are scaled with $\sqrt{2}$.
Finally, the \gls{svm} classifier predicts the \gls{mi} class.
%
%

We quantize the feature extraction to a mixture of \mbox{8-,} \mbox{16-,} and full precision 32-bit fixed- and floating-point representations and the \gls{svm} to 8-bit fixed-point, summarized in Fig.~\ref{fig:riemann:feature_extraction_quant}.
The decision on the precision depends on the trade-off between energy efficiency and accuracy preservation. With 8- or 16-bit fixed-point numbers, it is possible to exploit the \gls{simd} instructions. However, not all the parts of the \gls{mrc} can be quantized due to numerical instability and significant accuracy loss. 


\subsubsection{\gls{iir} Bandpass Filters}\label{ch:riemann:design:iir}

The input data $\mathbf{X} \in \mathbb{R}^{N_{ch} \times{N_{s}}}$ with dimensions number of \gls{eeg} channels $N_{ch}$ and number of time samples $N_s$, is quantized to 8 bits. 
Each channel is filtered with $f$ \gls{iir} bandpass filters. 
The filters can become unstable, especially with quantization.
The internal accumulators can diverge, even if the output remains bounded.
%
We implement the Direct-Form I defined in~\cite{smith2020filter},
since it does not experience numerical overflow in the internal signals, because all internal registers store either the input or the output of the filter~\cite{smith2020filter}.
%
A typical approach for quantizing an \gls{iir} filter is to express them as a cascade of \glspl{sos}, each of which can be quantized with different dynamic ranges, thus minimizing the effect of quantizing the filter coefficients on the impulse response. 
%
With 8-bit fixed-point quantization, the impact is significant, while with 12 bits these effects are minimal.
Therefore, we choose 12 bits for the filter coefficients to prevent overflows that would occur with 16 bits. 
We re-scale the intermediate results in between the \glspl{sos} to remain in the same dynamic range and accumulate them with 16-bit registers in order to use \gls{simd} operations for the following iteration. 
All dynamic ranges for all sections are chosen independently 
and forced to be a power of two to implement simple bit-shifts instead of expensive divisions.

\subsubsection{Covariance Matrix and Whitening} \label{ch:riemann:implementation:design:whitening}

Recall, that the covariance matrix $\mathbf{C} \in \mathbb{R}^{n \times n}$, in our case $n = N_{ch}$, including regularization, is computed as
\vspace{-0.2cm}
\begin{equation}\label{eq:riemann:design:covmat}
  \mathbf{C} = \mathbf{X} \mathbf{X}^T + \rho \mathbf{I}
  \vspace{-0.2cm}
\end{equation}
and Whitening is defined as
\vspace{-0.2cm}
\begin{equation}\label{eq:riemann:design:whitening}
  \mathbf{W} = \mathbf{C}_{\text{ref}}^{-1/2} \mathbf{C} \mathbf{C}_{\text{ref}}^{-1/2},
  \vspace{-0.2cm}
\end{equation}
with $\mathbf{C}_{\text{ref}}^{-1/2}$ being the reference matrix, computed by averaging the covariance matrices of all the training trials. For quantization, we define $n_c$ and $n_{\text{ref}}$ to be the number of bits to represent $\mathbf{C}$ and $\mathbf{C}_{\text{ref}}^{-1/2}$, respectively. 
Since we can exploit either 4- or 2-way \gls{simd} operations, we test both $n_c = n_{\text{ref}} = 8$ and $16$. However, the former yields a significant accuracy drop, while the latter causes overflows. Hence, we reduce $n_{\text{ref}}$, until training completes without overflow, resulting in $n_{\text{ref}} = 11$.
Our experiments have shown that using $n_c = 16$ and $n_{\text{ref}} = 11$ yields similar accuracy to the full-precision version.
Moreover, we force the scaling factor for the covariance matrix computation to be a power of two to exploit bit-shifts, while the dynamic range for the Whitening depends on the quantization of $\mathbf{C}$ and $\mathbf{C}_{\text{ref}}^{-1/2}$.
Finally, for the intermediate and final results of Eq.~\ref{eq:riemann:design:whitening}, we keep the full dynamic range with 32 bits since the input to the matrix logarithm is very sensitive to quantization errors, as explained next. 
%

\subsubsection{Matrix Logarithm}\label{ch:riemann:design:logm}

The matrix logarithm of a square, positive definite matrix $\mathbf{A} \in \mathbb{R}^{n \times n}$ is defined in terms of its \gls{evd}, as
\begin{equation}
  \text{logm}(\mathbf{A}) = \mathbf{Q}^{-1} \text{logm}(\mathbf{D}) \mathbf{Q},
\end{equation}
where $\mathbf{A} = \mathbf{Q}^{-1} \mathbf{D} \mathbf{Q}$, and the logarithm of a diagonal matrix $\mathbf{D}$ is computed by applying the logarithm to its diagonal elements. 
The whitened covariance matrix $\mathbf{W}$ in \gls{mrc} is dense and symmetric, allowing us to optimize the \gls{evd}. We first compute the tridiagonal decomposition to obtain a tridiagonal matrix $\mathbf{T}$ similar to the original one, i.e. the Eigenvalues are preserved. Then the \gls{evd} can be computed on $\mathbf{T}$ requiring less computational effort.
The final transformation is 
\begin{equation}
  \mathbf{W} = \mathbf{Q}_t^T \mathbf{T} \mathbf{Q}_t = \mathbf{Q}_t^T \mathbf{Q}_d^T \mathbf{D} \mathbf{Q}_d \mathbf{Q}_t,
\end{equation}
where $\mathbf{Q}_t$ is the orthogonal matrix for the tridiagonal transformation and $\mathbf{Q}_d$ the one for the \gls{evd}.
$\mathbf{Q}_d \mathbf{Q}_t$ is an orthogonal matrix containing the Eigenvectors of $\mathbf{W}$.
To compute the tridiagonal matrix, we use the Householder transformation~\cite{burden2004numerical}. The complexity of the transformation can be reduced by rearranging the operations and exploiting the sparsity of the vectors~\cite{burden2004numerical}.
%
For computing the diagonal matrix $\mathbf{D}$ from the tridiagonal symmetric matrix $\mathbf{T}$, we use the QR algorithm with implicit Wilkinson Shift~\cite{wilkinson1965evd}. 
The matrix logarithm only exists 
if the matrix is positive definite, 
meaning that all the Eigenvalues are positive. In full-precision \gls{mrc}, the input of the matrix logarithm is always positive definite, while with quantization the Eigenvalues change and in some cases even become negative, making it impossible to compute real logarithm. 
%
We address this issue by (a) making use of the entire 32-bit dynamic range for the inputs, and (b) clipping all Eigenvalues $\lambda_k$ to $\max\{\lambda_k, \lambda_{\min}\}$ by introducing a threshold $\lambda_{\min}=10^{-3}$ to ensure all Eigenvalues remain above zero. Its value is chosen based on the smallest Eigenvalue occurring while training the full precision \gls{mrc}. Moreover, both Householder transformation and QR algorithm are computed with 32-bit floating-point values. 
Finally, we convert the results back to 8-bit fixed-point format using the dynamic range learned during training. 

\subsubsection{\acrfull{svm}}
The final classifier in \gls{mrc} is a \gls{svm}, which we train on the quantized features.
The weights and biases are then quantized with bit-width $n_w = 8$ and $n_b = 32$, respectively, by determining the dynamic ranges after training. We do not rescale the output of the \gls{svm} because the prediction is made based on the relative largest output value. Hence, the weight vector can use the entire range available with 8 bit, reducing the quantization error.

\section{Implementation}\label{ch:riemann:implementation}

We implement the mixed-precision \gls{mrc} on \wolf{}~\cite{pullini2019wolf} which has a \gls{soc} domain and a compute cluster with 8 parallel \riscv{}-based processors called \riscy{}, or CV32E40P, implementing RV32IMFC \gls{isa} with custom \xpulp{} extensions for \gls{dsp}, e.g., SIMD instructions, hardware loops, post-incremental load and store~\cite{gautschi2017riscy}. The cluster cores have two shared \glspl{fpu} and 64\,kB of shared L1 memory via the \gls{tcdm} interconnect. More memory can be accessed via a \gls{dma} unit from the shared L2 memory (448\,kB) present in the \gls{soc} domain.

Our \gls{mrc} implementation is divided into three main blocks framed with blue, red, and green lines in Fig.~\ref{fig:background:riemannian}, respectively: (a) computation of the frequency bands until Whitening: each frequency band, highlighted with blue rectangle, is computed using 8 cores as described in the following paragraphs; (b) computation of the matrix logarithm and vectorization: every core computes one matrix logarithm followed by the vectorization concurrently with the other cores, i.e. 8 matrix logarithms, colored with red rectangle, are computed at the same time;
(c) \gls{svm} computed with a single core, colored in green.

\subsubsection{IIR Filter}\label{ch:impl:filter}

As described in Section~\ref{ch:riemann:design:iir}, we set the bit-width of the coefficients to $n_a = n_b = 12$, and the bit-width of the internal registers to $n_i = 16$.
Each \gls{sos} contains three \glspl{mac} for the forward accumulation 
and two \glspl{mac} for the backward accumulation. 
This enables the usage of \gls{simd} instructions with bit-width 16. 
%
We compute the filtered output of different \gls{eeg} channels on separate cores of the cluster to utilize the concurrent capabilities of \wolf{}.

\subsubsection{Covariance Matrix}\label{ch:impl:cov}

The computation of the covariance matrix is a \gls{mmm}, as shown in Eq.~\eqref{eq:riemann:design:covmat}, which results in a symmetric matrix.
Therefore, we only compute the upper right triangle and copy the remaining elements.
Since $\mathbf{X_k}$ is the filtered input data of band $k$, packed to 8 bits, the implementation makes use of \gls{simd} instructions to improve the performance significantly.
The computation is implemented concurrently by splitting the upper right part of the output matrix among all processing units.

\subsubsection{Whitening}\label{ch:impl:whiten}

Whitening consists of two \glspl{mmm}, as described in Eq.~\eqref{eq:riemann:design:whitening}.
Based on the quantization scheme described in Section~\ref{ch:riemann:implementation:design:whitening}, the first multiplication is computed in 16 bit, and the second in 32 bit.
For the first multiplication, we use 2-way \gls{simd} instructions.
We use the concurrent implementation found in the \gls{dsp} library~\cite{wang2019dsp} for \pulp, where each core computes a part of the matrix.

\subsubsection{Matrix Logarithm}

For computing the \gls{evd}, we implement both the basic version of Householder transformation and the improved version~\cite{burden2004numerical} for speedup analyses.
The computation of the rotation matrix required for the Givens rotation~\cite{givens1954numerical} of each QR step is done exclusively with multiplications, divisions, and additions, without using expensive trigonometric functions~\cite{bindel2002computing}.
%
For parallel implementation, every core is assigned with a frequency band and computes the Householder transformation and QR algorithm. 

\subsubsection{\acrfull{svm}}

The matrix-vector product of the \gls{svm} is computed using 8-bit \gls{simd} instructions.
We implement it on a single core, since it accounts for a negligible portion of the computation of the entire model.



\section{Experimental Results and Discussion}\label{ch:results}


We apply our methods on the BCI Competition IV-2a dataset~\cite{Brunner2008BCIA} with 22 \gls{eeg} channels and 4 \gls{mi} classes from 9 different subjects. 
There are 288 trials for each of the training and testing sets. Each trial lasts 6\,s and is sampled at 250\,Hz.

Table~\ref{tab:results:accuracy} reports the classification accuracy of our proposed models compared to related work with different \gls{mrc} configurations and \eegnet{}.
\gls{mrc} can be scaled to use more or fewer frequency bands and temporal windows.
Hersche et al.~\cite{Hersche2018FastFeatures} have shown that $f=43$ frequency bands and a single temporal window $t=1$ can already achieve comparable accuracy (74.8\% on average) to the full \gls{mrc} (75.5\%) while requiring $3\times$ fewer features.
In this work, we use only one temporal window $t=1$ of 3.5\,s and further scale down the number of frequency bands. Our results show that with $2.4\times$ less frequency bands, i.e. $f=18$, of bandwidth 2\,Hz between 4 and 40\,Hz, our full precision model achieves slightly higher accuracy by introducing the regularization with the hyperparameter $\rho=1$. 
%
Comparing to \eegnet{}, which is known to be a compact \gls{cnn} for \gls{bmi} applications~\cite{Lawhern2018EEGNet:Interfaces}, 
our full precision \gls{mrc} is 3.8\% more accurate.
%
%
Regarding the quantization, \eegnet{} can be quantized down to 8-bit precision for the entire network with Q-\eegnet{}~\cite{Schneider2020} without significant loss in accuracy (0.4\%). However, our proposed mixed-precision \gls{mrc} is still 3.2\% more accurate. The minimal loss in accuracy of 1\% from full to mixed-precision can be attributed mainly to the quantization at the input of the matrix logarithm. 
Regarding the memory footprint, Q-\eegnet{} requires 68.15\,kB, while our \gls{mrc} implementation uses approximately 84\,kB, i.e. 2$\cdot$22$\cdot$876 for 8-bit input and output of \gls{iir} filters, 18$\cdot$(22+1)$\cdot$22/2 for $W_k$ in 32-bits and reused for $L_k$, 18$\cdot$(22+1)$\cdot$22/2 for the model parameters $\mathbf{C}_{\text{ref,k}}^{-1/2}$ in 16 bits, and 4554$\cdot$4 for \gls{svm} weights in 8 bits. 



\setlength{\tabcolsep}{4.8pt}
\begin{table}[b]
\vspace{-0.4cm}
  \centering
  \caption{Classification accuracy (\%) on 4-class \gls{mi}.}\label{tab:results:accuracy}
  \vspace{-0.2cm}
  {\renewcommand{\arraystretch}{0.8}
    \small
    \begin{tabular}{lcccccc}
      \toprule
      & \multicolumn{2}{c}{Q-\eegnet{}} & \multicolumn{4}{c}{\gls{mrc}} \\
      \cmidrule(r){1-1}
      \cmidrule(r){2-3}
      \cmidrule(){4-7}
      Ref. & \cite{Schneider2020} & \cite{Schneider2020} & \cite{Hersche2018FastFeatures}$^\nmid$ & \cite{Hersche2018FastFeatures}$^\natural$ &  Ours$^\diamond$ & Ours$^\diamond$ \\
      Precision & full & 8-bit & full & full & full & mixed \\
      $t$\,/\,$f$\,/\,$\rho$ & & & 3\,/\,43\,/\,0 & 1\,/\,43\,/\,0 & 1\,/\,18\,/\,1 & 1\,/\,18\,/\,1 \\
      \cmidrule(r){1-1}
      \cmidrule(r){2-3}
      \cmidrule(){4-7}
      Subj. 1 & 81.0 & 81.0 & 90.0 & 91.8 & 91.8 & 90.7\\
      Subj. 2 & 57.6 & 53.1 & 55.5 & 51.6 & 53.7 & 51.2\\
      Subj. 3 & 87.9 & 91.2 & 81.3 & 83.5 & 83.5 & 81.0\\
      Subj. 4 & 61.6 & 58.1 & 71.9 & 73.3 & 73.7 & 74.1\\
      Subj. 5 & 70.6 & 68.4 & 69.6 & 63.4 & 68.8 & 63.0\\
      Subj. 6 & 53.4 & 50.1 & 56.7 & 58.6 & 56.7 & 56.3\\
      Subj. 7 & 75.7 & 75.2 & 85.6 & 86.7 & 84.1 & 58.9\\
      Subj. 8 & 77.4 & 81.2 & 83.8 & 81.6 & 81.5 & 82.7\\
      Subj. 9 & 76.7 & 79.7 & 84.9 & 82.6 & 82.2 & 81.8\\
      \cmidrule(r){1-1}
      \cmidrule(r){2-3}
      \cmidrule(){4-7}
      Avg. Acc. & 71.3 & 70.9 & 75.5 & 74.8 &  \textbf{75.1} & \textbf{74.1}\\
      Std. & 11.5 & 14.3 & 12.8 & 13.9 & 12.2 & 13.2\\
      \bottomrule
    \end{tabular}
  }
\end{table}

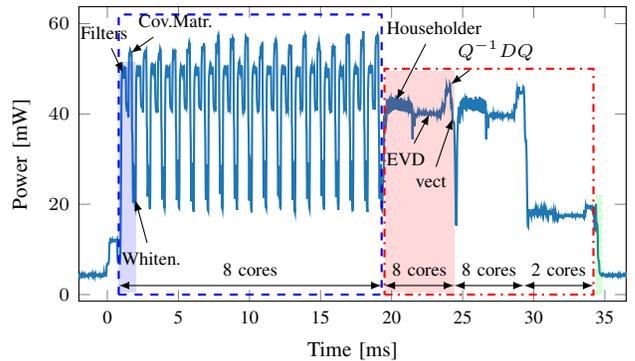
\begin{figure}[!t]
  \centering
  \begin{tikzpicture}
    \begin{axis}
    [
      no markers,
      width=\columnwidth,
      height=5.5cm,
      xlabel={Time [ms]},
      ylabel={Power [mW]},
      scaled ticks=false,
      xtick={0, 0.005, 0.01, 0.015, 0.02, 0.025, 0.03, 0.035},
      xticklabels={0, 5, 10, 15, 20, 25, 30, 35},
      xmin = -0.0020, xmax = 0.0365,
    ]
      \addplot+ [name path=f, thick] table [x=t, y=p, col sep=comma] {measurements/multiscale_prepared.csv};
      
      \draw[blue,thick,dashed] (0.0008,0) rectangle (0.0193, 62);
      \fill[blue,opacity=0.15] (0.0008,0.5) rectangle (0.002, 51.5);
      \draw[red,thick,dashdotted] (0.0195,0) rectangle (0.0342, 50);
      \fill[red,opacity=0.15] (0.0195,0.5) rectangle (0.0244, 49.5); 
      \draw[green,thick,dotted, fill=green, opacity=0.15] (0.0344,0) rectangle (0.0348, 22);
      
      \node[anchor=west] (sHT) at (axis cs: 0.019,59){\scriptsize Householder};
      \node (dHT) at (axis cs:0.020,41){};
      \draw[->](sHT)--(dHT);
      \node[anchor=east] (sSVD) at (axis cs: 0.023,30){\scriptsize EVD};
      \node (dSVD) at (axis cs:0.023,42){};
      \draw[->](sSVD)--(dSVD);
      \node[anchor=west] (sQDQ) at (axis cs: 0.024,54){\scriptsize $Q^{-1}DQ$};
      \node (dQDQ) at (axis cs:0.0235,45){};
      \draw[->](sQDQ)--(dQDQ);
      \node[anchor=east] (sVect) at (axis cs: 0.0245,25){\scriptsize vect};
      \node (dVect) at (axis cs:0.0245,41){};
      \draw[->](sVect)--(dVect);
      
      \node[] (sW) at (axis cs: 0.0031,8){\scriptsize Whiten.};
      \node (dW) at (axis cs:0.0018,23){};
      \draw[->](sW)--(dW);
      \node[anchor=east] (sF) at (axis cs: 0.0021,58){\scriptsize Filters};
      \node (dF) at (axis cs:0.0014,47){};
      \draw[->](sF)--(dF);
      \node[anchor=south] (sCM) at (axis cs: 0.0048,57){\scriptsize Cov.Matr.};
      \node (dCM) at (axis cs:0.001,52){};
      \draw[->](sCM)--(dCM);
      
      \draw[<->] (axis cs:0.0008,2) -- (axis cs:0.0193, 2) node [pos=0.5, anchor=south] {\scriptsize 8 cores};
      \draw[<->] (axis cs:0.0195,2) -- (axis cs:0.0244, 2) node [pos=0.5, anchor=south] {\scriptsize 8 cores};
      \draw[<->] (axis cs:0.0244,2) -- (axis cs:0.0293, 2) node [pos=0.5, anchor=south] {\scriptsize 8 cores};
      \draw[<->] (axis cs:0.0293,2) -- (axis cs:0.0342, 2) node [pos=0.5, anchor=south] {\scriptsize 2 cores}; 
    \end{axis}
  \end{tikzpicture}
  \vspace{-0.3cm}
  \caption{End-to-end power measurement. The colors match the compute blocks in Fig.~\ref{fig:background:riemannian} explained in Sec.~\ref{ch:riemann:implementation}.}
  \label{fig:results:power}
  \vspace{-0.3cm}
\end{figure}

\new{
Table~\ref{tab:results:riemann} shows the computation time and the performance impact of the optimizations and Fig.~\ref{fig:results:power} depicts the measured power trace.
The first 18 peaks are measured when the frequency bands are calculated using 8 cores, framed with blue dashed line. 
The \gls{iir} filter implementation achieves 3.77 \glspl{mac} per cycle with 7.26$\times$ parallel speedup.
Here, each output sample requires 10 \glspl{mac}, 3 shuffle operations, and 4 bit-shifts, resulting in a theoretical maximum of 5 \glspl{mac} per cycle.
The covariance matrix computation reaches 8.14 \glspl{mac} per cycle with concurrent execution yielding a speedup of 7.10$\times$ using 8 cores. 
The parallel speedup of the Whitening is 4.98$\times$ due to the parallelization overhead that is more visible with smaller matrix sizes (here 22$\times$22). However, it is not the bottleneck part of the \gls{mrc}.
%
The improvements of the Householder transformation have a significant impact on the performance yielding a speedup of 3.6$\times$ on the computation of the matrix logarithm compared to the baseline, while the parallel speedup is 5.67$\times$ compared to the single core computation and 20.64$\times$ compared to the baseline.
18 matrix logarithms are computed, distributed to the 8 cores on a first-come first-served schedule, i.e. twice 8 matrix logarithms are computed on 8 cores, then the remaining 2 on two cores, as reflected on the power trace, framed with red dashdotted line. This workload unbalance contributes negatively to the parallel speedup. However, the performance would not increase significantly with a more balanced distribution since the ideal speedup would be 6$\times$ with six parallel cores. 
Moreover, the maximal number of \glspl{flop} per cycle is 2, of which we reach 1.69, limited by the iteratively computed divisions and square root operations. 
Finally, the \gls{svm} accounts for a minimal part of the execution with 0.15\,ms, highlighted with green frame in Fig.~\ref{fig:results:power}.
}
For comparison, the embedded \gls{bmi} in~\cite{Belwafi2018AnSystems} consumes 0.7\,W and takes around 0.4\,s, more than an order of magnitude more in terms of both, power consumption and execution time---or two orders of magnitude worse in terms of energy efficiency.
We also compare to the Q-\eegnet{} implementation in~\cite{Schneider2020} that is publicly available. We run both Q-\eegnet{} and \gls{mrc} on \wolf{} at 100\,MHz and 1.1\,V. The former takes 13.64\,ms consuming 0.678\,mJ while the runtime of \gls{mrc} lays within the same order of magnitude with 33.39\,ms and consumes 1.304\,mJ. 
It is up to the user to decide on the trade-off between accuracy and cost depending on the application scenario.


\setlength{\tabcolsep}{3.9pt} 
\begin{table}[b]
\vspace{-0.4cm}
  \centering
  \caption{Computation time for \gls{mrc} on \wolf{} with a frequency of 100\,MHz at 1.1\,V.}\label{tab:results:riemann}
  \vspace{-0.2cm}
  {\renewcommand{\arraystretch}{0.8}
    \small
    \begin{tabular}{@{}lrrrrr@{}}
      \toprule
      & baseline & \makecell[r]{improved\\ \gls{evd}} & concurrent & \makecell[r]{parallel\\speedup} & ops/c$^\nmid$ \\
      \cmidrule(r){1-1}
      \cmidrule(r){2-4}
      \cmidrule(){5-6}
      Filter            & 66.67\,ms  & 66.67\,ms & 9.18\,ms & 7.26 & 3.77 \\
      Cov. matrix & 34.80\,ms  & 34.80\,ms & 4.90\,ms & 7.10 & 8.14 \\
      Whitening         & 24.29\,ms  & 24.29\,ms & 4.88\,ms & 4.98 & 0.79 \\
      Matrix logm.  & 309.76\,ms & 85.18\,ms & 15.01\,ms & 5.67 & 1.69 \\
      \gls{svm}         & 0.15\,ms   & 0.15\,ms  & 0.15\,ms & - & 1.25 \\
      \cmidrule(r){1-1}
      \cmidrule(r){2-4}
      \cmidrule(){5-6}
      Total             & 439.48\,ms & 206.93\,ms & \textbf{33.39\,ms} \\
      \glspl{mac}/cycle$^\natural$ & 0.62     & 0.62   & \textbf{4.11} \\
      \glspl{flop}/cycle$^\diamond$ & 0.08     & 0.30  & \textbf{1.69} \\
      insn/cycle        & 0.907    & 0.837 & 0.788\\
      \toprule
    \end{tabular}
  }
  \begin{minipage}{0.9\columnwidth}
    \scriptsize
    $^\natural$ Number of fixed-point \glspl{mac} over number of cycles w/o matrix logarithms. \\
    $^\diamond$ Number of \glspl{flop} over number of cycles during matrix logarithms. \\ 
    $^\nmid$ \glspl{mac} or \glspl{flop} per cycle for the concurrent implementation except \gls{svm}. 
  \end{minipage}
\end{table}

\section{Conclusion}\label{ch:conclusion}
This paper presents an improved \gls{mrc} with reduced model size while keeping comparable accuracy (75.1\% vs. 75.5\%~\cite{Hersche2018FastFeatures}), allowing accurate low-power embeeded \gls{bmi}. We further scale down the model by quantizing and proposing a mixed-precision implementation yielding a minimal accuracy loss of 1\%, which is still 3.2\% more accurate than the \gls{soa} embedded \gls{cnn} for \gls{bmi} named Q-\eegnet{}~\cite{Schneider2020}. We propose a parallel implementation on a low-power \gls{mcu} called \wolf{}, which takes only 33.39\,ms and consumes 1.304\,mJ. The higher accuracy compared to Q-\eegnet{} comes at the cost of a 2.4$\times$ longer execution time and a 1.9$\times$ higher energy consumption. However, it is still two orders of magnitude more energy efficient than other embedded solutions~\cite{Belwafi2018AnSystems}. We provide an insight on accuracy-cost trade-off for embedded \gls{bmi} models with actual implementation and measurements.
%



\clearpage
\bibliographystyle{IEEEtran}
\bibliography{bibliography,ref_michael_mendeley}

\end{document}